\documentstyle[prl,aps,graphicx,floats,epsf,color]{revtex}
\begin{document}
\baselineskip=12pt
\twocolumn[\hsize\textwidth\columnwidth\hsize\csname@twocolumnfalse\endcsname

\title
{Molecular Dynamics Simulation of Folding and Diffusion of Proteins in Nanopores}

\author
{Leili Javidpour,$^a$ Muhammad Sahimi,$^{b,\dagger}$ and M. Reza
Rahimi Tabar$^{a,c}$}

\vskip 1cm

\address
{$^a$Dep. of Physics, Sharif University of Technology, Tehran 11365, Iran\\
$^b$Mork Family Department of Chemical Engineering \& Materials
Science, University of Southern California, Los Angeles, California 90089-1211, USA\\
$^c$CNRS UMR 6529, Observatoire de la C$\hat o$te d'Azur, BP 4229,
06304 Nice Cedex 4, France}

\maketitle

%%%%%%%%%%%%%%%%%%%%%%%%%%%%%%%%%%%%%%%%%%%%%%%%%%%%%%
%ABSTRACT
%%%%%%%%%%%%%%%%%%%%%%%%%%%%%%%%%%%%%%%%%%%%%%%%%%%%%%

\begin{abstract}
A novel combination of discontinuous molecular dynamics and the
Langevin
equation, together with an intermediate-resolution model, are used to
carry out long (several $\mu$s) simulation and study folding transition
and
transport of proteins in slit nanopores. Both attractive ($U^+$) and
repulsive
($U^-$) interaction potentials between the proteins and the pore walls
are
considered. Near the folding temperature $T_f$ and in the presence of
$U^+$
the proteins undergo a repeating sequence of folding/partially-folding/
unfolding transitions, while $T_f$ decreases with decreasing pore
sizes. The
opposite is true when $U^-$ is present. The proteins' effective
diffusivity
$D$ is computed as a function of their length (number of the amino acid
groups), temperature $T$, the pore size, and the interaction potentials
$U^\pm$. Far from $T_f$, $D$ increases (roughly) linearly with $T$, but
due to
the thermal fluctuations and their effect on the proteins' structure
near
$T_f$, the dependence of $D$ on $T$ in this region is nonlinear. Under
certain conditions, transport of proteins in smaller pores can be {\it
faster}
than that in larger pores.


\end{abstract}
\hspace{.3in}
\newpage]

Proteins' importance to biological systems cannot be overstated
$\cite{1}$: as
enzymes they catalyze and regulate cells' activities; tissues are made
of
proteins, while as antibodies proteins are a vital part of the immune
system.
Proteins with globular structure fold into compact and biologically
active
configurations, and an important problem is understanding the
mechanisms by
which they attain their folded structure, and factors that contribute
to the
folding $\cite{2,3,4}$. Such understanding is important due to
debilitating
illnesses, such as Alzheimer's and Parkinson's diseases, that are
believed to
be the result of accumulation of toxic protein aggregates
$\cite{5,6,7,8}$, as
well as to the industrial production of enzymes and therapeutic
proteins based
on the DNA recombinant method $\cite{9}$.

While the three-dimensional (3D) structure of native proteins is
controlled by
their amino acid sequence $\cite{2,3,4}$, their transport properties
and the
kinetics of their folding depend on the local environment. But, whereas
protein
folding in dilute solutions under bulk condition, typically used in
{\it in
vitro} studies, is relatively well-understood, the more important
problem of
protein folding in a confined medium is not. The environment inside a
cell in
which proteins fold is crowded, with the volume fraction of the
crowding
agents (such as RNA) may be 0.2-0.3. Thus, even in the absence of
interactions
between proteins and other cellular molecules, their movement inside
the cell
is limited. The limitation affects proteins' stability. Experiments
indicated
$\cite{10}$ that confinement often stabilizes the proteins' native
structure
$\cite{11}$, denatures them in the limited space of the cage model,
first
suggested by Anfinsen $\cite{2,3,4}$, and {\it accelerates} folding
relative
to that in bulk solutions. Studies of proteins of different native
architectures in cylindrical nanpores indicated $\cite{12}$ that, {\it
in vivo}
folding is {\it not} always spontaneous; rather, a subset of proteins
may
require molecular chaperones.

Protein (enzyme) immobilization using porous solid support, via
adsorption,
encapsulation, and covalent linking, has been used for a long time
$\cite{13,14}$. Such practical applications as biocatalysis $\cite{15}$
and
biosensors also entail not only better understanding of the folding in
confined
media, but also transport of proteins in such media. At the same time,
protein purification using nanoporous membranes is also gaining
attention
$\cite{16}$. SiC nanoporous membranes $\cite{17}$ allow $\cite{18}$
diffusion
of proteins up to 29000 Daltons, but exclude larger ones. Despite the
fundamental and practical significance of transport of protein in
confined
media, there is currently little understanding of the phenomenon.

The goal of this paper is twofold. First, we use molecular dynamics
(MD)
simulation to study protein folding and stability in slit nanopores.
Second, we
utilize a novel combination of MD simulation and the Langevin equation
(LE) to
study protein transport in the nanopores. To our knowledge, our
combination of
the MD simulation and the LE has never been proposed before, nor has
there been
any simulation of transport of proteins in nanopores. For such
important
practical applications as membrane purification, biocatalysis, and
sensors, the
transport of proteins in nanopores is of utmost importance. A slit
nanopore is
a reasonable model for the type of pores that one encounters in such
applications $\cite{15,16,17,18}$ and, despite its simplicity, it might
also
be a reasonable model for the pores in biological membranes.

Some Monte Carlo $\cite{19}$ and MD $\cite{5,20}$ simulations of
proteins'
behavior in nanopores were reported before. In particular, Lu {\it et
al.}
$\cite{5}$ and Cheung {\it et al.} $\cite{20}$ studied folding of
proteins in
{\it spherical} pores of different radii. Cheung {\it et al.} studied
the
phenomenon as a function of the volume fraction of a crowding agent,
which
they modeled by a bed of hard spheres with repulsive interaction with
the
proteins. While a spherical pore may be a suitable model for the cavity
of
GroEl-GroES complex, it is not so for the pores of membranes,
biocatalysts, and
sensors that are of prime interest to us. Instead, the slit (and
cylindrical)
pores are more appropropriate. Moreover, for the types of applications
that we
consider, the pore space consists of interconnected channels, which is
completely different from what Refs. [5,20] considered.

In addition, the protein model that we use (see below) is, in our
opinion, much
more realistic than what the previous investigations $\cite{5,20}$
utilized.
For example, they used a simplified model for the amino acids that was
based on
two united atom (UA) beads. Moreover, the side chains of the amino-acid
residues were not explicitly considered. The model that we utilize
represents
the amino acids using four UA beads (see below), while the side chains
are also
considered explicitly, hence honoring the proteins' structure much more
realistically.

We simulate de novo-designed $\alpha$ family of proteins $\cite{21}$,
which
consists of only 4 types of amino acids in their 16-residue sequence,
simplified further $\cite{22}$ to a sequence of hydrophobic (H) and
polar (P)
residues. Using periodicity in the H-P sequence of the 16-residue
peptide
$\alpha_{1B}$, we made 3 other sequences with lengths $\ell=9$, 23 and
30
residues. As the four proteins have similar native structures, the
differences
in their behavior is attributed to their lengths. The simulations
indicated
that they all fold into an $\alpha$-helix.

The proteins are modeled by an intermediate-resolution model
$\cite{23,24,25,26}$, with several changes described below. Every amino
acid
is represented by four UA groups or beads. A nitrogen UA represents the
amide
N and hydrogen of an amino acid, a C$_\alpha$ UA represents the
$\alpha$-C
and its H, and a C UA for the carbonyl C and O. The fourth bead $R$
represents
the side chains, all of which are assumed to have the same diameter as
CH$_3$.
All the backbone bond lengths and bond angles are fixed at their ideal
values,
and the distance between consecutive C$_\alpha$ UA is fixed according
to
experimental data.

To carry out long and efficient simulations, we use discontinuous MD
(DMD)
$\cite{27}$. This allowed us to carry out 5 $\mu$s MD simulations, one
order
of magnitude longer than the previous simulations. The forces acting on
the
beads are the excluded-volume effect, and attraction between bonded and
pseudobonded beads, between pairs of backbone beads during HB
formation, and
between hydrophobic side chains. Nearest-neighbor beads along the chain
backbone are covalently bonded, as are the C$_\alpha$ and R UAs.
Pseudobonds
are between next-nearest neighbor beads along the backbone to keep the
backbone
angles fixed; between neighboring pairs of C$_\alpha$ beads to maintain
their
distances close to the experimental data, and between side chains and
backbone
N and C UAs to hold the side-chain beads fixed relative to the
backbone. All of
this keep the interpeptide group in the {\it trans} configuration, and
all the
residues as $L-$isomers, as required.

The potential $U_{ij}$ between a pair $ij$ of bonded beads, separated
by a
distance $r_{ij}$, is given by, $U_{ij}=\infty$, for, $r_{ij}\leq
l(1-\delta)$,
and $r_{ij}\geq l(1+\delta)$, and, $U_{ij}=0$ for
$l(1-\delta)<r_{ij}<l(1+
\delta)$. Here, $l$ is the ideal bond length, and $\delta=0.02375$ is
the
tolerance in the bond's length $\cite{23,24,25,26}$. The hydrophobic
(HP)
interactions between the side chains and the H in the sequence, if
there are at
least 3 intervening residues between them, is given by, $U_{\rm
HP}=\infty\;,
-\epsilon_{\rm HP},$ and 0 for, $r_{ij}\leq\sigma_{\rm HP}$,
$\sigma_{\rm HP}<
r_{ij}\leq 1.5\sigma_{\rm HP}$, and $r_{ij}>1.5\sigma_{\rm HP}$,
respectively, where $\sigma_{\rm HP}$ is the HP side-chains' diameter.

The HB interaction may occur between the N and C beads with at least 3
intervening residues, but each bead may not contribute to more than one
HB at any time, with the range of the interaction being about
$4.2\;\AA$. The
HBs are stable when the angles in N-H-O and C-O-H, controlled by a
repulsive
interaction between each of the N and C beads with the neighboring
beads of the other one, are almost $180^\circ$. Thus, if a HB is formed
between
beads N$_i$ and C$_j$, a repulsive interaction between neighbor beads
of N$_i$,
namely, C$_{i-1}$ and C$_{\alpha i}$, with C$_j$ is assumed, and
similarly for
the neighbor beads of C$_j$, namely, N$_{j+1}$ and C$_{\alpha j}$, with
the
N$_i$ bead.

An N or C bead at one end of the protein has only one neighbor bead in
its
backbone, instead of 2. Hence, controlling the HB angles will be
limited,
causing the HBs with one of their terminal constituents to be less
restricted
and, thus, more stable than the other HBs. This may cause formation of
non-$\alpha$-helical HBs in a part of the protein between the N and C
beads,
and of semistable structures that influence the results. To address
this
problem, assume that the N-terminal bead, N$_1$, has a HB with C$_j$.
For
$i=1$, bead C$_{i-1}$ does not exist to have a repulsive interaction
with C$_j$
and help control the HB angles. So, we use C$_{\alpha 1}$. Not only can
we
consider the repulsion between this bead and C$_j$, but also we define
an upper
limit for their distance so as to control the motion freedom of N$_1$
and C$_j$
that constitute the beads in the HB. The potential $U_{kl}$ of such
interactions is given by, $U_{kl}=\infty,\;\epsilon_{\rm HB},\; 0,$ and
$\infty$ for $r_{kl}\leq\frac{1}{2}(\sigma_k+\sigma_l)$, $\frac{1}{2}
(\sigma_k+\sigma_l)< r_{kl}\leq d_1$, $d_1<r_{kl}\leq d_2$, and,
$r_{kl}>d_2$,
respectively.

Two H atoms have chemical bonds with the nitrogen in the proteins'
N-terminal,
and are free to rotate around the N$_1$-C$_{\alpha 1}$ bond, while at
the same time satisfying the constraints on the angles between the
chemical
bonds of N$_1$. Thus, if a HB is formed, one of the two H atoms lies in
the
plane of N, O and C, such that the angles in N-H-O and C-O-H are as
close to
$180^\circ$ as possible. Hence, we force the maximum distance between
C$_{\alpha 1}$ and C$_j$ to be the same as the maximum distance $d_2$
between
C{$_{\alpha i}$} and C{$_j$} in the usual HBs, and similarly when the
C-terminal C$_\ell$ has a HB with N$_i$. This allows us to control the
angles
in a HB that contains N$_1$. The $T-$dependence ($T$ is dimensionless)
of $d_2$
(in $\AA$), obtained from separate MD simulations (the details will be
given
elsewhere), is, $d_2\simeq 5.53-0.019/T$ for N$_1$-C$_{\alpha j}$, and
$d_2\simeq 5.69-0.044/T$ for C$_\ell$-C$_{\alpha i}$.

There is also hard-core repulsion between two unbonded beads that have
no
HB and HP interactions. At the same time, interactions between a pair
of beads, separated along the chain by 3 or fewer bonds, are more
accurately
represented by those between the atoms themselves, not the UAs. Thus,
we
developed a variant of the previous models $\cite{23,24,25,26}$ to
account for
such interactions: the beads are allowed to overlap by up to 25\% of
their
bead diameters, while for those separated by 4 bead diameters the
allowed
overlap is 15\% of their diameters.

We use a slit nanopore, modeled as the space between two 2D
structureless
carbon walls in the $xy$ plane between $z=\pm h/2$, with periodic
boundary
conditions in the $x$ and $y$ directions. The interaction between the
walls and the protein beads is, $U_{\rm PW}=\infty,\;-\epsilon_{\rm
PW},\;0,\;
-\epsilon_{\rm PW},$ and $\infty$ for, $z_X\leq -(h/2-d_{3X})$,
$-(h/2-d_{3X})<
z_X\leq -(h/2-d_{3X}-d_{4X})$,
$-(h/2-d_{3X}-d_{4X})<z_X<h/2-d_{3X}-d_{4X}$,
$h/2-d_{3X}-d_{4X}<z_X\leq h/2-d_{3X}$, and $z_X\geq h/2-d_{3X}$,
respectively,
where $z_X$ is the distance between the center of a bead $X$ and the
walls. For
all the beads, $\epsilon_{\rm PW}=\frac{1}{8}\epsilon_{\rm HB}$, so
chosen to
represent realistically the competition between protein folding and its
beads'
interaction with the walls. To estimate $d_{3X}$ and $d_{4X}$, the
energy and
size parameters between the C atoms in the walls and various beads were
calculated using Lorentz-Berthelot mixing rules,
$\sigma_{CX}=\frac{1}{2}
(\sigma_C+\sigma_X)$, and $\epsilon_{CX}=\sqrt{\epsilon_C\epsilon_X}$,
where
$X=$N, C$_\alpha$, C and R. Then, using separate simulations (details
will be
given elsewhere), the interaction potential $U_{CX}$ between different
beads
was estimated. The distances at which U$_{CX}$ and its second
derivative were
zero were taken as $d_{3X}$ and $d_{3X}+d_{4X}$. The results (in $\AA$)
are,
$d_{3X}=2.85,\;3.02,\;3.14$, and 3.31, and $d_{4X}=0.96,\;1.01,\;0.98$,
and
1.12, for $X=$N, C$_\alpha$, C, and R, respectively.

We now describe how the effective diffusivity $D$ of the proteins is
computed.
To do so, one must take into account the effect of the solvent on the
motion
of the proteins. In the previous works the solvent's effect on the
proteins'
motion was included only implicitly by the HP attraction between the
side
chains. While this might be appropriate for studying the folding, it is
not so
for computing $D$, since the solvent's viscosity $\eta$ strongly and
directly
affects $D$. To explicitly include the solvent effect, we have
developed the
following model which, to our knowledge, is new.

We first carry out DMD simulation for a time period $\Delta t$. Suppose
that
the speeds of the proteins' center of mass (CM) at the beginning and
end of the
period $\Delta t$ are, respectively, ${\bf v}_b$ and ${\bf v}_e$. Since
the
solvent's viscosity affects the proteins' velocity in the pore, but the
time
scale over which this effect is important is much different from
$\Delta t$, we
apply, at the end of the time period $\Delta t$, the Langevin equation
(LE) to
the proteins' CM to correct their velocity due to the presence of the
solvent's
molecules. To do so, we represent a protein as a particle with a mass
$m$ and
an effective radius equal to its radius of gyration $R_g$. Then, the
force
{\bf F} on its CM is given by,
\begin{equation}
{\bf F}=m({\bf v}_e-{\bf v}_b)/\Delta t\;.
\end{equation}
The discretized LE is given by,
\begin{equation}
\Delta{\bf v}={\bf v}_n-{\bf v}_b={\bf F}\Delta t/m-\xi{\bf v}_b\Delta
t/m+
\Delta{\bf R}(\Delta t)\;,
\end{equation}
where, $\xi=6\pi R_g\eta$, $\Delta{\bf R}$ is a Gaussian random force
(with
zero mean and variance $2k_BT\xi\Delta t/m^2$), and ${\bf v}_n$ is the
speed
{\it after} applying the LE (acting as the ${\bf v}_b$ for the next LE
application). Thus,
\begin{equation}
d{\bf v}={\bf v}_n-{\bf v}_e=-\xi\Delta t{\bf v}_b/m+\Delta{\bf
R}(\Delta t)\;,
\end{equation}
which yields ${\bf v}_n$, the velocity of the proteins corrected for
the
solvent effect. The DMD simulation is then continued for another time
period
$\Delta t$ using ${\bf v}_n$ as the ${\bf v}_b$, the LE is applied
again to
correct the proteins' velocity at the end of the period, and so on.

We now present the results of our simulations. Consider the case of
attractive
interaction potential $U^+$ between the proteins and the pore's walls.
A folded
state attaches itself to the walls only through its end groups, while
unfolded
ones may completely attach themselves to the walls. Thus, with $U^+$
the
decrease in the average potential energy of the unfolded states is
larger than
the corresponding decrease for the folded one. Hence, compared to the
bulk, the
unfolded states in the pore with a $U^+$ are more stable than the
folded one,
in qualitative agreement with Refs. [4,20] for spherical cavities.

Figure 1 shows a sequence of events for a protein of size $\ell=16$ in
a pore
of size $h=1.75$ nm at $T_f\simeq 0.13$. The protein changes its state
from
completely folded to a partially folded to an unfolded one which is
completely
attached to the pore's wall (frame D). Due to $U^+$, the transitions
occur
easily and repeatedly, even after a long time. Note that, in moving
from B to
C, the set of deformed $\alpha-$helical HBs changes, hence indicating
rapid
dynamics of the HB formation and deformation near $T_f$.

Also shown in Figure 1 are protein configurations (frames E and F with
pore
size $h=1.5$ nm) for $\ell=9$ at $T=0.08$. In these pores, a protein of
length
$\ell=9$ does not attain its native state at low $T$. Instead, it has a
U shape
with its two sides attached to the walls; it has 4 HBs, only one of
which is
$\alpha-$helical (the native state has 5 $\alpha-$helical HBs), and
more of its
atoms are close to the walls than those in the folded state. Although
the
potential energy of such unfolded states is roughly the same as one in
the
folded one, entropic effects which favor the unfolded states are also
important. Upon further cooling at $T$ below the apparent folding
temperature
$T_f$, the protein becomes trapped in the U shape without enough
kinetic energy
to overcome the energy barrier to attain a folded state. Thus, such
configurations do not represent truly folded states.

We also find that $T_f$ decreases with the pore size, which is due to
the
attractive interaction energy between the protein and the walls. The
decrease
in $T_f$ is indicative of the more stable unfolded states (or less
stable
folded state).

The opposite (namely, increasing $T_f$ with decreasing pore size) is
true if
the interaction is purely repulsive, $U^-$. In that case, proteins are
even
more confined in the pores. Thus, the number of possible unfolded
states and,
hence, their total entropy, will be smaller $\cite{28}$. But, due to
its
compact configuration, confinement affects the entropy of the folded
state less
strongly, implying that, in the presence of $U^-$ the folded state is
more
stable than that in the bulk.

Figure 2 presents the average interaction energy $\langle U_{\rm
PW}\rangle$
of the proteins with the walls, computed by the weighted histogram
analysis
method $\cite{29}$. Cooling the proteins at $T>T_f$ increases $|\langle
U_{\rm PW}\rangle|$, as well as the average $\langle n_\alpha\rangle$
of the
$\alpha-$helical HBs (which is, however, very small). Near T$_f$
$\langle
n_\alpha\rangle$ is nonnegligible, and the proteins can only laterally
attach
themselves to the walls, hence decreasing $|\langle U_{\rm
PW}\rangle|$. By
lowering $T$ further, nearly the entire $\alpha$-helix is formed, and
$|\langle U_{\rm PW}\rangle|$ increases again. Thus, Figure 2 indicates
that,
not only does the interaction with a nanopore disturb folding, but also
folding
to a definite structure disturbs the proteins' interaction with the
pore.

Before describing the results for the effective diffusivities, we
should point
out that, over the temperature range that we have simulated, the
proteins
resemble a prolate ellipsoid. Thus, they become elongated in very small
nanopores, while they can ``stand up'' in larger pores, such that their
ends
touch the pores' walls, if they are long enough. Such configurations
are also
shown in Figure 1. Thus, transport in small pores may actually be
faster than
in larger pores, a counterintuitive, but important result. Such
phenomena can
complicate further delineation of the dependence of $D$ on the various
important parameters of the system.

Our simulations indicate that transport of the proteins in the $xy$
planes
(parallel to the pore's walls) is Fickian, so that, after a
sufficiently long
time (which depends on the proteins' length, the pore size, and
$U^\pm$) the
mean-square displacements (MSDs) of the proteins' CM vary linearly with
the
time, hence yielding an effective diffusivity $D$ for the proteins. In
contrast, the MSDs in the direction perpendicular to the pore's walls
saturate
at a finite time.

Figure 3 presents the proteins' effective diffusivity $D$ in the $xy$
planes,
in a pore with fixed $h/R_g=4.1$. As $T$ increases and $T_f$ is
approached,
$D$ also increases, since the proteins' $R_g$ decreases. Far from $T_f$
(where
$R_g$ changes little) $D$ varies roughly linearly with $T$. However,
near
$T_f$, there are strong thermal fluctuations due to which $R_g$
strongly varies
with $T$. Consequently, $D$ no longer varies linearly with $T$.
Increasing the
proteins' length decreases $D$, as expected.

Figure 4 presents the diffusivities for a protein of fixed length,
$\ell=16$,
in four different pores, and compares the results with those in the
bulk. Note
that, the $T_f$ for the bulk state and those for the pores are not the
same.
Therefore, due to the strong fluctuations of $R_g$ and $D$ in the
region around
$T_f$, the pore and bulk diffusivities cannot be directly compared. The
comparison is possible mainly at temperatures far from $T_f$ (both
below
and above $T_f$). As Figure 4 indicates, at least for the $D_{\rm
bulk}\geq
D_{\rm pore}$ (taking into account the numerical uncertainties.

The results shown in Figures 3 and 4 were for an attractive potential
$U^+$
between the proteins and the pores' walls. Figure 5 compares the
diffusivities
for a protein of length $\ell=16$ in a pore of size $h=1.75$ nm for
both
attractive and repulsive interaction potentionals $U^\pm$, and compares
them
with the bulk values. Generally speaking, for $T\leq T_f$ the
diffusivities
with $U^-$ are larger than those with $U^+$.

In our current work, we are carrying out extensive simulations for
computing
the effective diffusivity $D$ as a function of the strength of the
interaction
potential $U_{\rm PW}$ between the proteins and the walls, its sign
(attractive
versus repulsive), and other relevant parameters. The results will be
presented
in a future paper.

We thank M. R. Ejtehadi, C. K. Hall, and M. D. Niry for useful
discussions. The
work of LJ was supported by Iranian Nanotechnology Initiative.

\newpage
\cleardoublepage

%FIGURE 1
\begin{figure}
\begin{center}
\begin{minipage}[t]{0.95\textwidth}
\centering
\includegraphics[width=0.72\textwidth]{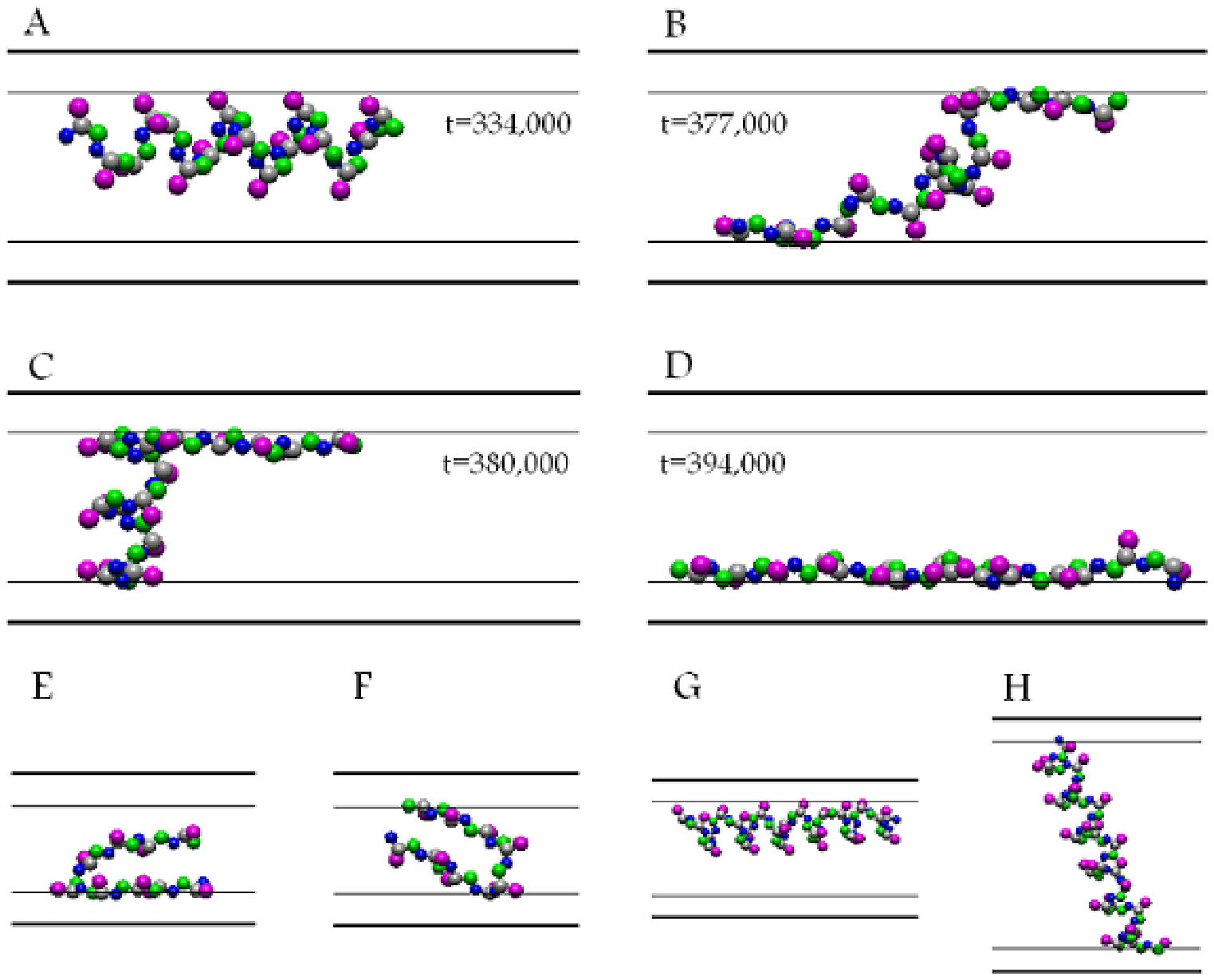}
\renewcommand{\figurename}{FIG.}
\caption{\small Various configurations (A-D) that a protein of length
$\ell=16$
takes on in a pore at $T_f\simeq 0.13$. Blue, gray, green, and pink
spheres
show, respectively, the N bead, C$_\alpha$, C, and the side chains. In
between
the walls and the thin lines at a distance $d_3$, $U_{\rm PW}=\infty$,
beyond
which the $U^+$ acts for a distance $d_4$. Times $t$ are in ps. Frames
E and F
show a short protein (lipid) with $\ell=9$ that has not reached its
native
state. Frames G and H show the configurations of a protein parallel and
perpendicular to the walls.}
\label{fig1}
\end{minipage}
\end{center}
\end{figure}

\newpage
\cleardoublepage

%FIGURE 2
\begin{figure}
\begin{center}
\begin{minipage}[t]{0.95\textwidth}
\centering
\includegraphics[width=0.72\textwidth]{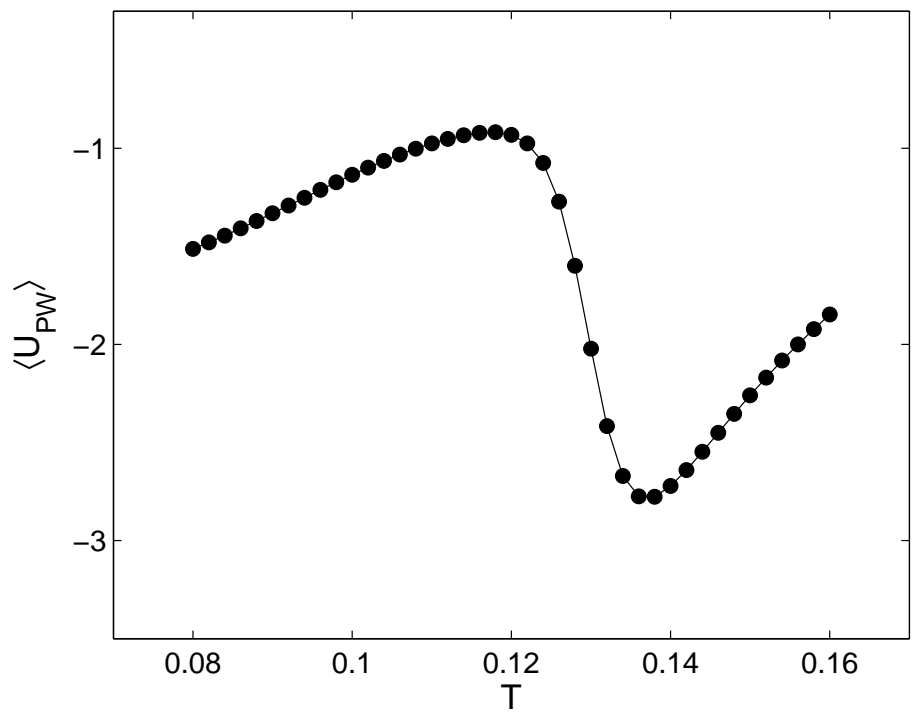}
\renewcommand{\figurename}{FIG.}
\caption{\small The average interaction energy $\langle U_{\rm
PW}\rangle$ for
a pore of size $h=1.75$ nm and protein of length $\ell=16$.}
\label{fig2}
\end{minipage}
\end{center}
\end{figure}

\newpage
\cleardoublepage

%FIGURE 3
\begin{figure}
\begin{center}
\begin{minipage}[t]{0.95\textwidth}
\centering
\includegraphics[width=0.72\textwidth]{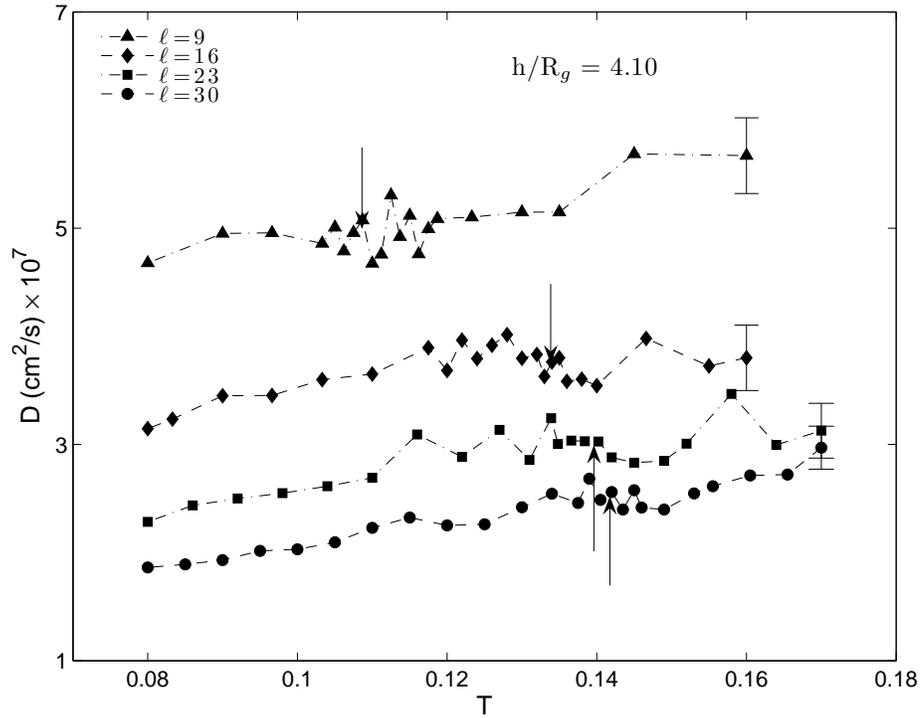}
\renewcommand{\figurename}{FIG.}
\caption{\small Dependence of the effective diffusivity $D$ on the
temperature
$T$ for various protein length $\ell$, in a pore with $h/R_g=4.1$.
Arrows
indicate the location of the folding temperature $T_f$.}
\label{fig3}
\end{minipage}
\end{center}
\end{figure}

\newpage
\cleardoublepage

%FIGURE 4
\begin{figure}
\begin{center}
\begin{minipage}[t]{0.95\textwidth}
\centering
\includegraphics[width=0.72\textwidth]{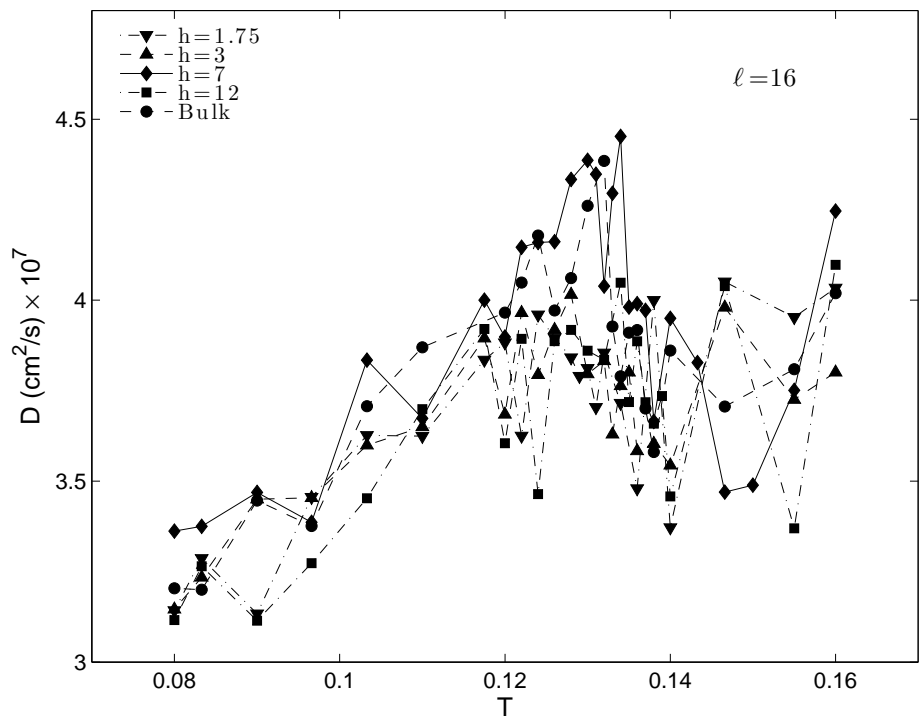}
\renewcommand{\figurename}{FIG.}
\caption{\small Dependence of the effective diffusivity $D$, for a
protein of
size $\ell=16$, on the temperature $T$ and the pore size $h$. The
radius of
gyration $R_g$ is constant, computed at $T=0.08$ under bulk condition
($R_g/h=0$).}
\label{fig4}
\end{minipage}
\end{center}
\end{figure}

\newpage
\cleardoublepage

%FIGURE 5
\begin{figure}[h]
\begin{center}
\begin{minipage}[t]{0.95\textwidth}
\centering
\includegraphics[width=0.72\textwidth]{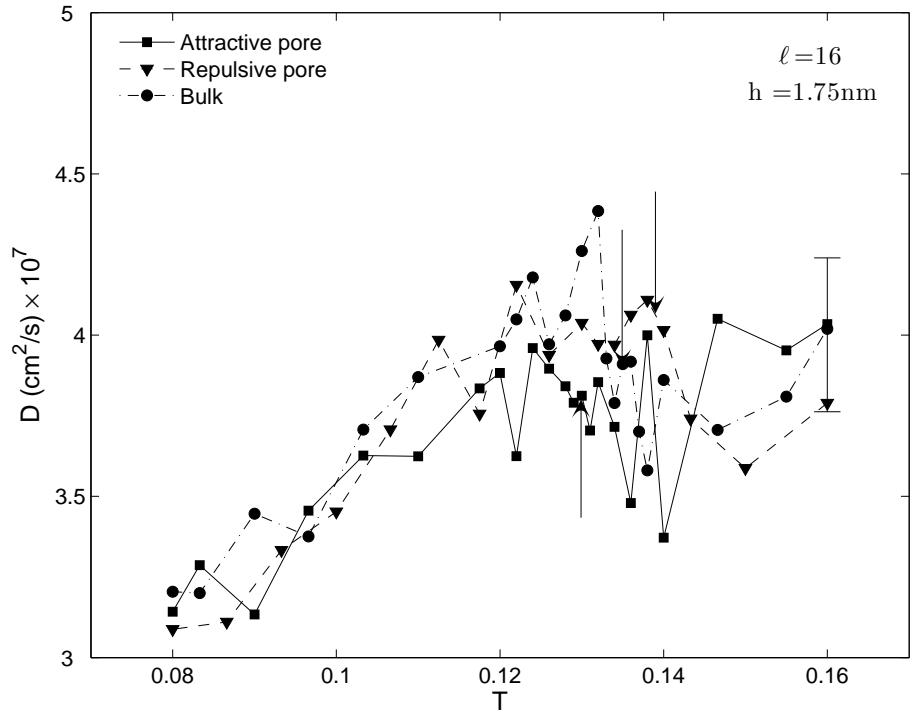}
\renewcommand{\figurename}{FIG.}
\caption{\small Comparison of the effective diffusivities in a pore
with
attractive and repulsive interactions between the protens and the
pore's
walls. The arrows indicate the location of $T_f$.}
\label{fig5}
\end{minipage}
\end{center}
\end{figure}

\end{document}